\documentclass[aps,prl,twocolumn,floats,amsmath,amssymb,showpacs,floatfix]%
{revtex4}
\usepackage{graphicx}
\usepackage{dcolumn}

\begin{document}

\title{Variable-range projection model for turbulence-driven collisions}
\author{K. Gustavsson$^{1}$, B. Mehlig$^{1}$, M. Wilkinson$^{2}$ and V. Uski$^{2}$}
\affiliation{
$^{1}$Department of Physics, G\"oteborg University, 41296
Gothenburg, Sweden \\ $^{2}$ Department of Mathematics and Statistics,
The Open University, Walton Hall,
Milton Keynes, MK7 6AA, England\\}

\begin{abstract}
We discuss the probability distribution of relative speed
$\Delta V$ of inertial particles suspended in a highly
turbulent gas when the Stokes numbers, a dimensionless measure
of their inertia, is large. We identify a mechanism giving rise to
the distribution $P(\Delta V) \sim  \exp(-C |\Delta V|^{4/3})$ (for some constant $C$).
Our conclusions are supported by numerical simulations and
the analytical solution of a model equation of motion.
The results determine the rate of collisions between suspended particles.
They are relevant to the hypothesised mechanism for formation of planets by  aggregation of dust particles in circumstellar nebula.
\end{abstract}


\maketitle
{\sl 1. Introduction}. It is widely believed that the first
stage of the formation of planets involves the aggregation
of microscopic dust grains in the gaseous nebula around
young stars \cite{Gol73}. This process must occur in a turbulent environment,
because the transport of angular momentum by diffusion would be too slow to account for the lifetimes of
these nebula. Also, the aggregation process occurs in gas with a
very low density, so that the motion of the dust grains is very lightly
damped. It is necessary to achieve a good understanding
of the relative velocity of collisions of the dust grains to
determine  whether and how planet formation could result from
the aggregation of microscopic dust grains. The relative velocity
is required to determine the rate of collision of the dust grains.
Also, if the relative velocity is sufficiently high, clusters may fragment upon collision. These issues concerning planet formation are discussed in \cite{Bec00,Wil08}.

Earlier discussions of the relative velocity of suspended particles \cite{Abr75,Vol80,Meh07} have estimated the order of magnitude of the relative velocity, but a satisfactory theory for its distribution has been lacking. In the context of planet formation, the case of lightly damped particles is most important. If the microscopic correlation time of the flow is $\tau$ and the damping rate (defined by (\ref{eq: 2}) below) is $\gamma$, we define the Stokes number as ${\rm St}=1/\gamma \tau$. A theoretical approach is required, because simulations are impracticable for the lightly damped case where ${\rm St}\gg 1$.

\begin{figure}[t]
\includegraphics[width=8cm]{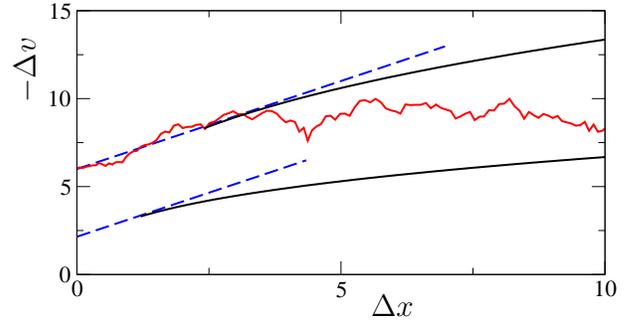}
\caption{\label{fig: 1} Variable-range projection model.
We show curves of constant probability $\rho(\Delta x,\Delta v)$
for $\Delta x \gg 1$ (black), $\Delta v \propto \Delta x^{1/3}$. Also
shown is a realisation of a trajectory of equation (\ref{eq: 6}) projected from large separations to $\Delta x=0$ (red), compared with
optimal trajectories (blue dashed). Here $\Delta x$ and $\Delta v$ are
dimensionless
variables in equation (\ref{eq: 6}).}
\end{figure}

In this letter we show that the probability distribution
function for the relative velocities $\Delta V$ of colliding particles is well approximated by
\begin{equation}
\label{eq: 1}
P(\Delta V)=A \exp\left(-{\cal C}\vert\Delta V\vert^{4/3}\gamma^{2/3}/{{\cal E}}^{2/3}
\right)
\end{equation}
where ${\cal E}$ is the turbulence intensity (the rate of dissipation per unit mass) and ${\cal C}$ is a universal dimensionless constant (with $A$ determined by normalising the distribution). We argue that this is a precise asymptote for the distribution for large  $\vert \Delta V\vert$.

We remark that there are connections with the distribution of accelerations in turbulent flows. The acceleration of a suspended particle is proportional to its velocity relative to the fluid. Because the relative velocity of two particles with ${\rm St}\gg 1$ is the sum of their (statistically independent) velocities relative to the fluid, the tail of the distribution of accelerations $a$ of suspended particles is of the form $P(a)\sim \exp[-{\rm const}|a|^{4/3}]$, analogous to (\ref{eq: 1}). For suspended particles with ${\rm St\ll 1}$, the acceleration is the same as Langrangian fluid acceleration, which also has a distribution of the same form as 
(\ref{eq: 1}), with $4/3$ replaced by $\approx 2/5$ \cite{LaP01}. The distribution of accelerations for suspended particles in a turbulent flow was studied numerically for a range of values of ${\rm St}$ by Bec {\sl et al} \cite{Bec06}. The results (figure 2b of their paper) are compatible with the limiting cases discussed above.

Our explanation of the  mechanism underlying equation (\ref{eq: 1}) proceeds as follows. The colliding particles acquire a relative velocity
when they are accelerated by different regions of the fluid. They are then \lq projected' (i.e. thrown) a certain distance away
from the fluid element which accelerated them. Since relative particle
velocities imparted by the fluid flow increase with separation, particles
which collide with a high relative velocity acquired their relative motion
when their separation was large. Our estimate of the probability distribution
function $P(\Delta V)$ involves a maximisation of the probability
of reaching zero separation with respect to variation of the distance
over which the particles are projected by the flow. We term this model
the \lq variable range projection' model. It has much in common with
the \lq variable-range hopping' model for electrical conduction in
semiconductors at low temperatures \cite{Mot68}, which also arises from
an optimisation of the hopping length and leads to an expression for the conductance of the form (\ref{eq: 1}), with temperature playing the role
of the relative velocity.
\begin{figure}
\includegraphics[width=8cm]{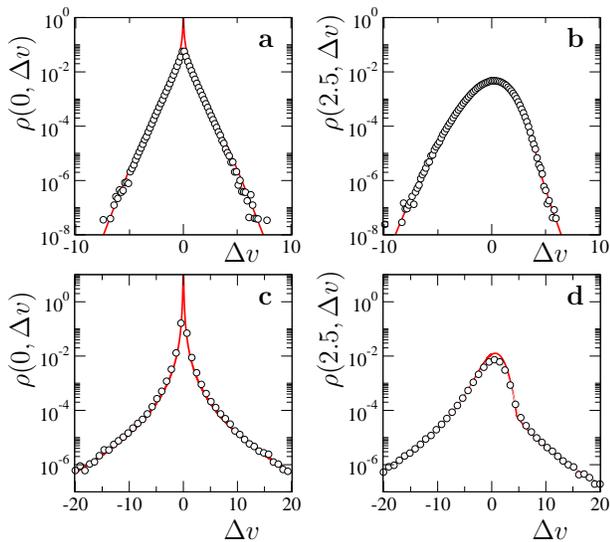}
\caption{\label{fig:2}
Probability density $\rho(\Delta x,\Delta v)$ for a simulation of equations (\ref{eq: 6}), (\ref{eq: 7}) ($\circ$), compared with theory (solid line): {\bf a} $\Delta x = 0$, compared with (\ref{eq: 1}). {\bf b} $\Delta x = 2.5$, compared with distribution obtained from (\ref{eq: 9},\ref{eq: 10},\ref{eq: 11}).
In both cases $\alpha = 2/3$ and $\epsilon=1$. The empirical distributions are normalised and the prefactor of the theoretical curves chosen to give the best fit. Panels {\bf c} and {\bf d} are the same as for {\bf a} and {\bf b} respectively, except $\alpha=4/3$.}
\end{figure}

Our heuristic description is supported by precise asymptotic
analysis of a one-dimensional model, equation (\ref{eq: 6}) below.
Figure \ref{fig:2}{\bf a} shows a comparison with simulation. For non-zero separation, the relative velocity distribution has a more complex asymmetric
form, Fig. \ref{fig:2}{\bf b}. We also confirm a surmise about the variance of the relative velocity \cite{Meh07}.

{\sl 2. Equations of motion}.
The equations of motion for the position $\mbox{\boldmath$x$}$ and velocity
$\mbox{\boldmath$v$}$ of a suspended particle are
$\dot{\mbox{\boldmath$r$}}=\mbox{\boldmath$v$}$ and
$\dot{\mbox{\boldmath$v$}}=\gamma [\mbox{\boldmath$u$}(\mbox{\boldmath$r$},t)-\mbox{\boldmath$v$}]\,,$
where $\mbox{\boldmath$u$}(\mbox{\boldmath$r$},t)$ is the fluid
velocity. This equation is applicable even
when the gas mean free path is large compared to the size of the particles \cite{Eps24}.
The corresponding equation for the relative
displacement $\Delta \mbox{\boldmath$X$}$ and relative velocity $\Delta \mbox{\boldmath$V$}$ of two particles is
\begin{equation}
\label{eq: 2}
\Delta \dot{\mbox{\boldmath$X$}}=\Delta \mbox{\boldmath $V$}\ ,\ \ \
\Delta \dot{\mbox{\boldmath$V$}}=\gamma [
\Delta {\mbox{\boldmath$u$}}(\Delta {\mbox{\boldmath$X$}},t)
-\Delta \mbox{\boldmath$V$}]
\end{equation}
and where $\Delta {\mbox{\boldmath$u$}} =
\mbox{\boldmath{$u$}}(\Delta{\mbox{\boldmath$X$}},t)
-{\mbox{\boldmath$u$}}(0,t)$.
According to the Kolmogorov theory of turbulence, there is a
range of lengthscales $\ell$ for which a component $\Delta u$ of the relative velocity of fluid elements with separation $\ell$ is determined only by the turbulence intensity. Dimensional arguments \cite{Fri97} then imply
\begin{equation}
\label{eq: 3}
\langle \Delta u(\ell,t)\Delta u(\ell,0)\rangle=({\cal E}\ell)^{2/3}
f(t{\cal E}^{1/3}/\ell^{2/3})
\end{equation}
for some function $f$
(angular brackets are used to denote averages throughout this paper).

{\sl 3. Variable-range projection model}.
Consider the relative displacement $\Delta X$ and speed $\Delta V$ of two
particles. When $\Delta X$ is small, the driving effect of the fluid
velocity $\Delta u$ is negligible, and the damping term is most significant. At greater distances, the relative velocity of
the background fluid drives the relative motion of the particles.
First let us consider the relative motion in greater detail at small separations,
such that we can neglect the effect of the driving term $\Delta u$. In this case $\Delta V$ decays exponentially in time, so that if two particles collide
with relative velocity $\Delta V$ at time $t$, their relative velocity at
an earlier time $t_0$ was $\Delta V_0(t_0)=\Delta V\exp[\gamma(t-t_0))]$. Integrating
this expression, we find that the relative separation at time $t_0$ was
\begin{equation}
\label{eq: 4}
\Delta X_0(t_0)=\int_t^{t_0}\!\!\!\!{\rm d}t'\ \Delta V {\rm e}^{\gamma(t-t')}
=\frac{\Delta V}{\gamma}\left(1-{\rm e}^{-\gamma(t-t_0)}\right)
\end{equation}
so that $\Delta V_0=\Delta V-\gamma \Delta X_0$, where $\Delta X_0$ was the initial separation.
Continuing to neglect the effects of the fluid velocity, we see that
in order for particles to collide with relative velocity $\Delta V$,
they must have had a larger velocity difference $\Delta V_0$ at a larger,
and unknown, separation $\Delta X_0$. For large $\Delta \mbox{\boldmath$X$}$, equations (\ref{eq: 2}) resemble those of an Ornstein-Uhlenbeck process \cite{Orn30}, where the velocity is Gaussian distributed. We therefore expect that for sufficiently large $\Delta X_0$, the relative velocity is
approximately Gaussian distributed:
\begin{equation}
\label{eq: 5}
\rho(\Delta V_0,\Delta X_0)\sim\frac{1}{\sqrt{2\pi \langle \Delta V_0^2\rangle}}
\exp\left[-\left(\frac{\Delta V_0^2}{2\langle \Delta V_0^2\rangle}\right)\right]\ .
\end{equation}
Here we use the expectation that for large separations, the relative velocity
is well approximated by the relative velocity of the fluid elements, so that
equation (\ref{eq: 3}) implies that
$\langle \Delta V_0^2\rangle\sim ({\cal E}\Delta X_0)^{2/3}$.
To determine where the inbound particle colliding with relative
velocity $\Delta V$ originated, we therefore find the value of the
separation $\Delta X_0$ which maximises the probability of colliding
with relative velocity $\Delta V$, that is we maximise $\rho(\Delta V_0,\Delta X_0)$,
where $\Delta V_0=\Delta V-\gamma \Delta X_0$, with respect to $\Delta X_0$.
Figure \ref{fig: 1} illustrates the trajectories.
Let the value for which the maximum obtains be $\Delta X_0^\ast$.
Neglecting the pre-exponential factor of (\ref{eq: 5}), we find
$\Delta X_0^\ast=-\Delta V/{2\gamma}$.
The distribution of velocities for colliding particles is predicted
to be $P(\Delta V)=\rho(\Delta V-\gamma \Delta X_0^\ast, \Delta X^\ast)$.
Neglecting the pre-exponential factor, we obtain
equation (\ref{eq: 1}). For the variance of the relative
velocity, it follows that $\langle \Delta V^2\rangle\propto {\cal E}/{\gamma}$. This provides a justification for
a result which was previously inferred from the Kolmogorov theory of
turbulence by a dimensional argument \cite{Meh07}.

{\sl 4. Microscopic model}.
The motion of the smallest eddies in a fully-developed turbulent flow
are characterised by the Kolmogorov length $\eta$, Kolmogorov time $\tau$ and Kolmogorov velocity $u_{\rm K}$. The flow is characterised by two dimensionless variables, the Stokes number, ${\rm St} = 1/{\rm \gamma \tau}$, and the Kubo number, ${\rm Ku}=u_{\rm K}\tau/\eta$. In turbulent velocity fields $\eta$, $\tau$ and $u_{\rm K}$ are functions of the dissipation rate ${\cal E}$ and the kinematic viscosity $\nu$. Dimensional considerations then imply that ${\rm Ku}=O(1)$. However in the following we consider a model for a turbulent flow in which ${\rm Ku}\ll 1$, corresponding to a very rapidly fluctuating flow field, which can be modelled by a Langevin equation.

Consider the equations of motion (\ref{eq: 2}) in one spatial dimension. We convert to dimensionless variables, writing $t'=\gamma t$, $\Delta x = \Delta X/\eta$, $\Delta v =\Delta V/\eta\gamma $.
When the velocity field $\Delta \mbox{\boldmath$u$}$ is very rapidly fluctuating, we can approximate the equation of motion in scaled variables by the following Langevin equation
\begin{equation}
\label{eq: 6}
{\rm d} \Delta x=\Delta v\,{\rm d}t'\,, \ \
{\rm d} \Delta v\!=\!-\Delta v{\rm d}t'+ \delta w
\end{equation}
where the random increment $\delta w$ satisfies
\begin{equation}
\label{eq: 7}
\langle \delta w \rangle =0\,, \ \
\langle \delta w^2\rangle \!= \!2 {\cal D}(\Delta x) {\rm d} t'
\,,\ \
{\cal D}(\Delta x)=\epsilon \vert \Delta x\vert^\alpha \,.
\end{equation}
Here we have introduced a parameter $\epsilon\sim {\rm Ku}^2$. Having
approximated equations (\ref{eq: 2}) by equations (\ref{eq: 6}), we find that solutions of (\ref{eq: 6}) for different values of $\epsilon$ can be obtained from the solution with $\epsilon=1$ by a scaling transformation. Although it suffices to consider the case where $\epsilon=1$, we retain $\epsilon$ in subsequent expressions because it will be used as a small parameter of a WKB expansion. This formal procedure allows us to study the tails of the joint probability distribution of $\Delta x$ and $\Delta v$ in a controlled manner. In (\ref{eq: 7}) we also allow for an arbitrary exponent $0 \leq \alpha < 2$. The value of $\alpha$ is determined by requiring that the variance of the relative velocity has the correct behaviour as $\Delta x\to \infty$: the solution of (\ref{eq: 6}) presented below indicates that $\langle \Delta v^2\rangle \sim \vert\Delta x\vert^\alpha$ for $\Delta x\to \infty$, so comparison with (\ref{eq: 3}) indicates that $\alpha=2/3$ is the correct choice.

{\sl 5. Distribution of collision velocities}.
The distribution (\ref{eq: 1}) of collision velocities
is determined by the joint distribution
$\rho(\Delta x,\Delta v)$ evaluated at $\Delta x = 0$.
To determine  $\rho(\Delta x, \Delta v)$
we solve the steady-state Fokker-Planck equation corresponding
to equations (\ref{eq: 6}), (\ref{eq: 7}):
\begin{equation}
\label{eq: 8}
0=-\Delta v\, \partial_{\Delta x} \,\rho+ \partial _{\Delta v}
\bigl(\Delta v\,\rho\bigr)
+ \epsilon |\Delta x|^\alpha\,\partial^2_{\Delta v}\,\rho\ .
\end{equation}
We note that at large values of $\Delta x$ ($\Delta x\gg \Delta v$),
the distribution
$\rho(\Delta x,\Delta v)$ is
Gaussian in $\Delta v$ [see equation (\ref{eq: 5})].
In order to solve (\ref{eq: 8}) we make a WKB ansatz \cite{Fre84}
\begin{equation}
\label{eq: 9}
\rho(\Delta x,\Delta v)=
K(\Delta x,\Delta v)\, \exp[-S(\Delta x, \Delta v)/\epsilon]\,.
\end{equation}
We write
\begin{eqnarray}
\label{eq: 10}
S(\Delta x,\Delta v)&=&
|\Delta x|^{2-\alpha} \tilde g_0(z,\Delta x)\,\\
\label{eq: 11}
K(\Delta x,\Delta v)&=&\exp[ -\tilde g_1(z,\Delta x)]\,,
\end{eqnarray}
where $z = s_1 \Delta v/\Delta x$ ($s_1 = \pm 1$ is chosen so that
$z > 0$).  Assuming that $\tilde g_0(\Delta x,z)$
does not depend on $\Delta x$, substituting (\ref{eq: 10}),
(\ref{eq: 11}) into (\ref{eq: 8}), and collecting terms in $\epsilon^{-1}$,
we obtain
\begin{equation}
\label{eq: 12}
g_0'(z) = \frac{z (s_1+z)+s_2 \sqrt{z^2 (z+s_1)^2-4g_0(z) z s_1(2-\alpha)}}{2 s_1}
\end{equation}
where $s_2 = \pm 1$ labels which branch of the square root is to be chosen.
In the following we label the solutions of (\ref{eq: 12})
by $g_0^{(s_1,s_2)}(z)$. Which of the solutions must be picked is determined by the boundary conditions.

Let us first consider an initial condition $(\Delta x,\Delta v)$ with a
positive and large value of $\Delta x$. Since $z >0$ by definition, $s_1$ determines
the sign of $\Delta v$. At large values of $\Delta x$ we know
that the distribution of $\Delta v$ is Gaussian
[eq. (\ref{eq: 5})]. This determines the small-$z$ asymptote of $g_0$:
$S=\Delta v^2/(2|\Delta x|^\alpha)
=|\Delta x|^{2-\alpha}z^2/2$. Thus we must require
$g_0 \sim z^2/2 \quad \mbox{as}\quad z \rightarrow 0$.
We find that only the solutions $g_0^{(-,-)}$ and $g_0^{(+,+)}$ match
this boundary condition.
In order
to reach $\Delta x=0$ from $\Delta x > 0$ the initial relative velocity
must be negative. For $\Delta x>0$ we are thus forced to choose $s_1 = -1$,
that is to consider the branch $g_0^{(-,-)}$.
Consider the case depicted in Fig. \ref{fig: 1} of
a particle projected to $\Delta x = 0$ determining
the distribution of collision velocities.
The action is determined by the large-$z$ behaviour of $g_0^{(-,-)}$, that is
$S = \lim_{\Delta x\rightarrow 0}|\Delta x|^{2-\alpha}
g_0^{(-,-)}(-\Delta v/\Delta x)$. We find
$g_0^{(-,-)}(z)\sim a_0(\alpha)\,z^{2-\alpha}\quad\mbox{for large $z$}$.
The prefactor $a_0(\alpha)$ is determined
by numerical integration. We find
$a_0(2/3) \approx 0.870$. The resulting action at $\Delta x=0$ is
\begin{equation}
\label{eq: 13}
S(\Delta x=0,\Delta v)=a_0(\alpha)\Delta v^{2-\alpha}\,.
\end{equation}
To determine the prefactor consider terms of order $\epsilon^0$ arising
from substituting (\ref{eq: 10}), (\ref{eq: 11}) into (\ref{eq: 8}):
\begin{equation}
\label{eq: 14}
0=g_0''-1-s_1 x z \partial_x \tilde g_1+(z+s_1z^2-2g_0')\,
\partial_z\tilde g_1
\end{equation}
We make the following separation ansatz
$\tilde g_1(x,z) = \lambda \log \Delta x + g_1(z)$.
It is motivated by the fact that it allows us to match
$\rho(\Delta x,\Delta v)$ to the known behaviour  (\ref{eq: 5})
at large separations. Inserting this ansatz
into (\ref{eq: 14}) we obtain (neglecting a normalisation constant)
\begin{eqnarray}
\label{eq: 15}
\tilde g_1 = \lambda \log\Delta x+ \int_{z_0}^z {\rm d}z'
\frac{1-g_0''(z') + s_1 \lambda
z'}{z'+s_1{z'}^2-2g_0'(z')}\,.
\end{eqnarray}
Consider now the limiting form of the prefactor
$K$ for large and for small separations $\Delta x$.
First, the limit of large $\Delta x$ corresponds to the limit
$z \rightarrow 0$. In this limit
$g_1$ is constant and to match the prefactor to the
known behaviour (\ref{eq: 5})
we must set $\lambda = 3\alpha/2$.
Second, the limit of $\Delta x\rightarrow 0$ corresponds
to the limit of $z \rightarrow \infty$. In this
limit the integrand in (\ref{eq: 15}) behaves as
$\sim {\lambda}/{z'} = {3\alpha}/({2 z'})$.
Integrating over $z$ we find that
${\rm e}^{-\tilde g_1} = |\Delta v|^{-3\alpha/2}$.
The final result (neglecting a normalisation factor) is thus
\begin{equation}
\label{eq: 16}
\rho(0,\Delta v)
= |\Delta v|^{-3\alpha/2}\,
\exp\big[-{\epsilon}^{-1} a_0(\alpha)|\Delta v|^{2-\alpha}
\big]\,.
\end{equation}
This result, for $\alpha = 2/3$, corresponds to the distribution
(\ref{eq: 1}) predicted by the variable-range projection model.
But here it has been derived, including the algebraic prefactor,
from a microscopic model. Fig. \ref{fig:2} {\bf a}, {\bf c}
compares of (\ref{eq: 16}) with simulations of the Langevin equation (\ref{eq: 6}).

{\sl 6. Relative velocities at larger separations}.
For non-zero separations, our WKB approximation is complicated by the
fact that different branches, corresponding to different
choices signs $s_1$, $s_2$ in (\ref{eq: 12}), must be combined.
For each branch, at finite values of $\Delta x$, the contribution
to $\rho(\Delta x, \Delta v)$ is of the form (\ref{eq: 9}),
with the action given by (\ref{eq: 10}) and (\ref{eq: 12}), with the prefactor given by eqs. (\ref{eq: 11}) and (\ref{eq: 15}). Which branches must be chosen depends upon the signs of $\Delta x$ and $\Delta v$. If two branches contribute for given values of $\Delta x$ and $\Delta v$, the branch with the smallest action dominates. The branches which are available correspond to four different choices of signs in the construction of solutions of (\ref{eq: 12}), namely $g_0^{(s_1,s_2)}(z)$. We already noted that only the solutions $g_0^{(-,-)}(z)$ and $g_0^{(+,+)}(z)$ can match the correct asymptotic behaviour at small $z$, namely $g_0\sim z^2/2$.

Let us consider the case where $\Delta x>0$. When $\Delta v<0$ (that is, when $s_1=-1$), we find that only the branch with action determined by the function $g_0^{(-,-)}(z)$ contributes, with corresponding action
\begin{equation}
\label{eq: 17}
S(\Delta x,\Delta v)=
|\Delta x|^{2-\alpha} g_0^{(-,-)}(- \Delta v/\Delta x)\,.
\end{equation}
This expression tends to (\ref{eq: 13}) as $\Delta v \rightarrow -\infty$,
and to the Gaussian form $S(\Delta x,\Delta v) \sim \epsilon^{-1}{\Delta v^2}/{|\Delta x|^\alpha}$ for small values of $\Delta v$.

For $\Delta v>0$ however, the WKB solution is more
complicated. For small $z$, and for sufficiently small $\Delta v$ the solution is given by the branch $g_0^{(+,+)}(z)$. This solution increases very rapidly as $z$ increases; we find $g_0^{(+,+)}(z)\sim (1+\alpha)z^3/9$ as $z\to \infty$, so this branch of the WKB solution becomes very small for large
$\Delta v$. By adapting the argument in section 3 above, however, we can argue that the tails of the probability density for the velocity should in fact be given by a branch where the action is $S\sim a_0(\alpha)\Delta v^{2-\alpha}$ for $\Delta v\to\infty$, where the prefactor $a_0(\alpha)$ is the same as for the $\Delta v<0$ branch. It is possible to find a solution for the branch $g_0^{(+,-)}(z)$ with the correct behaviour, namely
$g_0^{(+,-)}(z)\sim a_0(\alpha)\,z^{2-\alpha}$ as $z\to \infty$. This condition also ensures that
the tails of $\rho(\Delta x, \Delta v)$ are consistent
with (\ref{eq: 13})  in the limit $\Delta x\rightarrow 0$.

For $\Delta v>0$, we therefore construct the solution using two branches. For $0\le z\le z^\ast$ the solution constructed from $g_0^{(+,+)}(z)$, satisfying the $g_0^{(+,+)}(z)\sim z^2/2$ for $z\to 0$, is dominant. For $z>z^\ast$, the solution constructed from $g_0^{(+,-)}(z)$, satisfying $g_0^{(+,-)}(z)\sim a_0(\alpha)\,z^{2-\alpha}$ dominates. The point $z^\ast$ is determined by the condition that the action of the two solutions is equal, that is $g_0^{(+,+)}(z^\ast)=g_0^{(+,-)}(z^\ast)$. We remark that the solution $g_0^{(+,-)}(z)$ only exists for $z>z_{\rm c}$, where $z_{\rm c}$ is the critical point at which the discriminant in (\ref{eq: 12}) vanishes.
Fortunately, we find $z^\ast>z_{\rm c}$ (for $\alpha = 2/3$ we find $z_{\rm c}\approx 0.14$). The prefactor is given by eqs. (\ref{eq: 11}) and  (\ref{eq: 15}). Figure \ref{fig:2}{\bf b}, {\bf d} compares our distribution with simulations for $\Delta x\ne 0$.

{\sl 7. Conclusions}. In this letter we have shown how the distribution of relative velocities of particles suspended in highly turbulent flow at large ${\rm St}=1/(\gamma\tau)$ may be surmised from an optimisation argument which we term \lq variable range projection', leading to equation (\ref{eq: 1}). We validated this simple and general heuristic argument by a WKB analysis of a one-dimensional Langevin equation model, which produces an identical relative velocity distribution at zero separation.

{\sl Acknowledgements.} We acknowledge discussions with J. Bec and support from Vetenskapsr\aa{}det and from  the research initiative
\lq Nanoparticles in an interactive
environment' at G\"oteborg university.


\begin{thebibliography}{0}

\bibitem{Gol73} P. Goldreich and W. R. Ward,
{\it Astrophys. J.}, {\bf 183}, 1051-61, (1973).

\bibitem{Bec00}
S. V. W. Beckwith, T. Henning and Y. Nakagawa, Dust properties and assembly
of large particles in protoplanetary disks, in {\sl Protostars and protoplanets IV},
eds. V. Manning, A. P. Boss and S. Russell, University of Arizona Press, (2000).

\bibitem{Wil08}
M. Wilkinson, B. Mehlig and V. Uski,
{\it Astrophys. J. Suppl.}, in press - arXiv: astro-ph/0706.3536.

\bibitem{Abr75}
J. Abrahamson,
{\it Chem. Eng. Sci.}, {\bf 30}, 1371-9, (1975).

\bibitem{Vol80}
H. J. V\"olk, F. C. Jones, G. E. Morfill and S.  R\"oser,
{\it Astron. \& Astrophys.}, {\bf 85}, 316, (1980).

\bibitem{Meh07}
B. Mehlig, V. Uski and M. Wilkinson,
{\it Phys. Fluids}, {\bf 19}, 098107, (2007).

\bibitem{LaP01} A. La Porta, G. A. Vith, A. M. Crawford, J. Alexander,
and E. Bodenschatz, {\sl Nature}, {\bf 409}, 1017, (2001).

\bibitem{Bec06} J. Bec, L. Biferale, G. Boffetta, A. Celani, M. Cecini,
A. Lanotte, S. Musacchio and F. Toschi,
{\it J. Fluid. Mech.}, {\bf 550}, 349, (2006).

\bibitem{Mot68}  N. F. Mott, {\sl J. Non-Cryst. Solids}, {\bf 1}, 1,
(1968).

\bibitem{Eps24} P. S. Epstein,
{\it Phys. Rev.}, {\bf 23}, 710, (1924).

\bibitem{Fri97} U. Frisch, {\sl Turbulence}, Cambridge University Press,
(1997).

\bibitem{Orn30}
G. E. Ornstein and L. S. Uhlenbeck, {\it  Phys. Rev.}, {\bf 36}, 823,
(1930).

\bibitem{Fre84} M. I. Freidlin and A. D. Wentzell, {\sl Random Perturbations
of Dynamical Systems}, Springer: New York (1984).

\end{thebibliography}
\end{document}